\documentclass[a4paper]{article}

\usepackage{INTERSPEECH2020}
\usepackage{cite}
\usepackage{url}
\usepackage{hyperref}
\usepackage{verbatim}
\usepackage[table,xcdraw]{xcolor}
\usepackage{array}
\usepackage{tabularx}
\newcolumntype{Y}{>{\centering\arraybackslash}X}

\title{Quasi-Periodic Parallel WaveGAN Vocoder: A Non-autoregressive Pitch- dependent Dilated Convolution Model for Parametric Speech Generation}
\name{Yi-Chiao Wu$^{1}$, Tomoki Hayashi$^1$, Takuma Okamoto$^2$, Hisashi Kawai$^2$, and Tomoki Toda$^{12}$}

\address{
  $^1$Nagoya University, Japan\\
  $^2$National Institute of Infomation and Communications Technology, Japan
 }
\email{yichiao.wu@g.sp.m.is.nagoya-u.ac.jp, tomoki@icts.nagoya-u.ac.jp}

\begin{document}
\maketitle
\begin{abstract}
In this paper, we propose a parallel WaveGAN (PWG)-like neural vocoder with a quasi-periodic (QP) architecture to improve the pitch controllability of PWG. PWG is a compact non-autoregressive (non-AR) speech generation model, whose generative speed is much faster than real time. While utilizing PWG as a vocoder to generate speech on the basis of acoustic features such as spectral and prosodic features, PWG generates high-fidelity speech. However, when the input acoustic features include unseen pitches, the pitch accuracy of PWG-generated speech degrades because of the fixed and generic network of PWG without prior knowledge of speech periodicity. The proposed QPPWG adopts a pitch-dependent dilated convolution network (PDCNN) module, which introduces the pitch information into PWG via the dynamically changed network architecture, to improve the pitch controllability and speech modeling capability of vanilla PWG. Both objective and subjective evaluation results show the higher pitch accuracy and comparable speech quality of QPPWG-generated speech when the QPPWG model size is only 70~\% of that of vanilla PWG.
\end{abstract}

\section{Introduction}

Because of the high temporal resolution of speech signals, speech waveform modeling is challenging. The general technique to tackle speech synthesis (SS) is called a vocoder~\cite{vocoder_1939,vocoder_1966}, which encodes speech into low-dimensional acoustic features and decodes speech on the basis of these acoustic features. The conventional vocoders such as STRAIGHT~\cite{straight} and WORLD (WD)~\cite{world} are usually designed on the basis of a source-filter model~\cite{source_filter}, which decomposes speech into spectral and prosodic acoustic features. However, the ad~hoc signal-processing mechanisms imposed on the conventional vocoders cause the loss of phase information and temporal details, which results in marked speech quality degradation.

To achieve high fidelity SS, many neural network (NN)-based autoregressive (AR) SS models such as SampleRNN~\cite{samplernn} and WaveNet (WN)~\cite{wavenet} have been proposed to directly model the probability distributions of speech waveforms without many ad~hoc assumptions of SS. The NN-based vocoders~\cite{sd_wn_vocoder,si_wn_vocoder,ns_wn_vocoder,srnn_vocoder} are also proposed on the basis of these AR SS models to replace the synthesizers of the conventional vocoders for recovering the lost phase information and temporal details and generating high-quality speech. Furthermore, because of the extremely slow generation of WN and SampleRNN, many AR models with compact networks and specific knowledge~\cite{fftnet,wavernn,lpcnet} and non-AR models such as flow-based~\cite{pwn,clarinet,waveglow,flowavenet,waveffjord} and generative adversarial network (GAN)~\cite{gan}-based models~\cite{pwg,melgan, gantts} have been proposed for real-time speech generation.

However, because of the data-driven nature and the lack of prior speech knowledge, it is hard for these NN-based SS models to deal with unseen data. For instance, if the pitches of testing acoustic features are scaled or outside the observed pitch range of training data, the pitch accuracy and speech quality of the WN-generated speech samples markedly degrade. Since the pitch controllability is an essential feature for a vocoder, NN-based SS models with carefully designed periodic and aperiodic inputs~\cite{nsf_2019, nsf_2020, pap_gan} greatly improve the pitch modeling capability. Furthermore, in our previous work, we proposed a quasi-periodic WN vocoder (QPNet)~\cite{qpnet_2019,qpnet_2020}, which adopts pitch-dependent dilated convolution networks (PDCNNs) to dynamically change the network architecture according to the input pitches, to improve the pitch controllability and speech modeling ability of WN.

To tackle the slow generation of AR WN/QPNet, we apply a quasi-periodic (QP) structure to a compact non-AR model parallel WaveGAN (PWG)~\cite{pwg}. Since PWG transforms a noise sequence sampled from a standard Gaussian distribution into speech samples by taking conventional acoustic features as the auxiliary feature, PWG is more flexible than the models required specific periodic and aperiodic inputs. Moreover, the non-AR fashion also makes the parallelized generation avaliable for real-time generation. In this paper, we propose a fast and flexible QPPWG vocoder to improve the pitch controllability and speech modeling efficiency of PWG. Both objective and subjective evaluations are conducted, and the experimental results show the higher pitch accuracy, comparable speech quality, and smaller model size of the QPPWG vocoder than that of the PWG vocoder.

\section{Parallel WaveGAN}

\begin{figure}[t]
\begin{center}
\includegraphics[width=0.77\columnwidth]{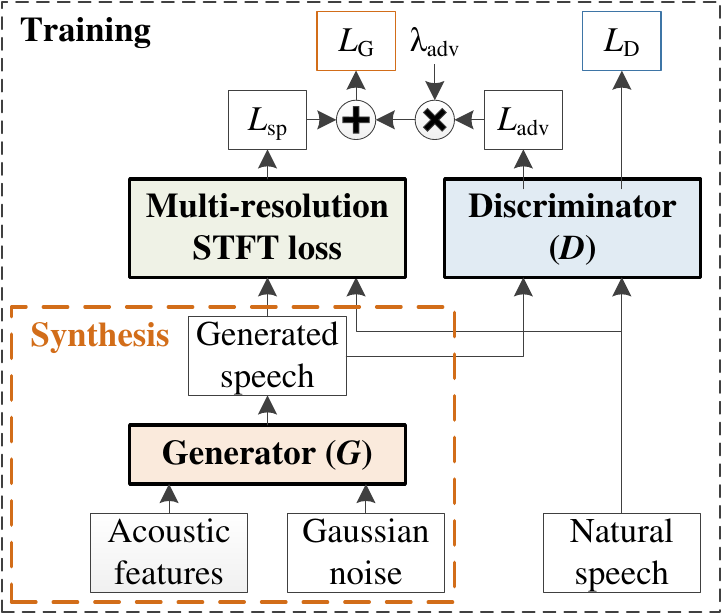}
\vspace{-2mm}
\caption{Architecture of Parallel WaveGAN.}
\label{fig:pwg}
\end{center}
\vspace{-10mm}
\end{figure}

As shown in Fig.~\ref{fig:pwg}, PWG is composed of a discriminator ($D$), a generator ($G$), and a multi-resolution short-time Fourier transform (STFT) loss module. The discriminator is trained to detect synthesized samples as fake speech and natural samples as real speech. The training criterion of the discriminator is to minimize the loss $L_{D}$, which is formulated as
\begin{align}
&L_{\mathrm{D}}(G, D) \nonumber \\
&=\mathbb{E}_{\boldsymbol{x} \in p_{\mathrm{data}}}\left[(1-D(\boldsymbol{x}))^{2}\right]+\mathbb{E}_{\boldsymbol{z} \in N(0, I)}\left[D(G(\boldsymbol{z}))^{2}\right],
\end{align}
where $\boldsymbol{x}$ denotes the natural samples, $p_{\mathrm{data}}$ denotes the data distribution of the natural samples, $N(0, I)$ denotes a Gaussian distribution with zero mean and standard deviation, and
$\boldsymbol{z}$ denotes the input noise of the generator drawn from the Gaussian distribution. Note that all auxiliary features of $G$ are omitted in this chapter for simplicity. The discriminator is a fully-convolutional network, which consists of several dilated convolution network (DCNN)~\cite{dcnn} layers with LeakyReLU~\cite{leakyrelu} activation functions. The dilation size of each DCNN layer is extending in an exponential growth manner with a base of two and an exponent of its layer index.  The generator is trained to generate speech samples, which makes the discriminator difficult to distinguish between the synthesized and natural samples. The training criterion of the generator is to minimize the generator loss ($L_{\mathrm{G}}$) formulated as
\begin{align}
L_{\mathrm{G}}(G, D)=L_{\mathrm{sp}}(G)+\lambda_{\mathrm{adv}} L_{\mathrm{adv}}(G, D),
\end{align}
which is the weighted sum of an adversarial loss ($ L_{\mathrm{adv}}$) from the GAN structure and a spectral loss ($L_{\mathrm{sp}}$) from the multi-resolution STFT loss module with a weight $\lambda_{\mathrm{adv}}$. The $L_{\mathrm{adv}}$ is formulated as 
\begin{align}
L_{\mathrm{adv}}(G, D)=\mathbb{E}_{\boldsymbol{z} \in N(0, I)}\left[(1-D(G(\boldsymbol{z})))^{2}\right].
\end{align}
Unlike flow-based models~\cite{pwn,clarinet,waveglow,flowavenet,waveffjord} adopting an invertible network to transform the distribution of the input noise to the real data distribution, PWG adopts a GAN structure to make the generator learn the transformation via the feedback from the discriminator. The generator adopts a WN-like network but without the AR mechanism and causality, so the generation of PWG markedly faster than that of WN.  To ensure the stability and efficiency of the GAN training, PWG adopts an extra $L_{\mathrm{sp}}$ loss as a regularizer of the generator. Specifically, the $L_{\mathrm{sp}}$ is a summation of a spectral convergence loss ($L_{\mathrm{sc}}$) and a log STFT magnitude loss ($L_{\mathrm{m}}$). The $L_{\mathrm{sp}}$ is formulated as
\begin{align}
L_{\mathrm{sc}}(\boldsymbol{x}, \hat{\boldsymbol{x}})&=\frac{\||\mathrm{STFT}(\boldsymbol{x})|-|\mathrm{STFT}(\hat{\boldsymbol{x}})|\|_{F}}{\|\mathrm{STFT}(\boldsymbol{x}) |\|_{F}},
\end{align}
and the $L_{\mathrm{m}}$ is formulated as
\begin{align}
L_{\mathrm{m}}(\boldsymbol{x}, \hat{\boldsymbol{x}})&=\frac{1}{N}\|\log |\mathrm{STFT}(\boldsymbol{x})|-\log |\mathrm{STFT}(\hat{\boldsymbol{x}})|\|_{L1},
\end{align}
where $\hat{\boldsymbol{x}}$ denotes the generated samples from the generator, $\left \| \cdot  \right \|_{F}$ is Frobenius norm, $\left \| \cdot  \right \|_{L1}$ is $L1$ norm, $\left | \mathrm{STFT}\left ( \cdot  \right ) \right |$ denotes STFT magnitudes, and ${N}$ is the number of the magnitude elements. Moreover, to avoid the generator overfitting to a specific STFT resolution causing a suboptimal problem, the final $L_{\mathrm{sp}}$ is a summation of several $L_{\mathrm{sp}}$ values calculated on the basis of STFT features with different analysis parameters such as FFT size and frame length and shift.

Although PWG achieves high fidelity speech generation, the fixed and generic network architecture of PWG is not appropriate. Speech is a quasi-periodic signal consisting of periodic components with long-term correlations and aperiodic components with short-term correlations, so the fixed architecture of PWG modeling all components is inefficient. Specifically, a {\it receptive field} is a region in the input space that the outputs are affected by, and the {\it receptive field} length is highly related to the modeling capacity. However, modeling speech using fixed-length {\it receptive fields} may lead to the {\it receptive fields} including many redundant samples from the oversampled periodic components. To tackle this problem, we proposed a QPPWG vocoder to respectively model periodic and aperiodic components using cascaded fixed and pitch-adaptive modules.

\section{Quasi-Periodic parallel WaveGAN}

\begin{figure}[t]
\includegraphics[width=1.0\columnwidth]{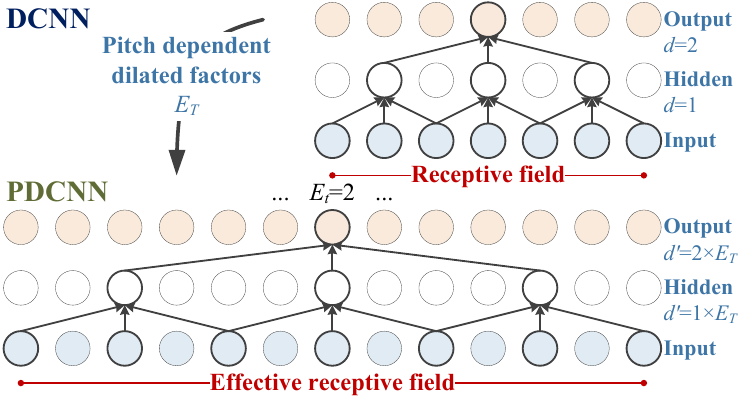}
\vspace{-5mm}
\caption{Pitch-dependent dilated convolution.}
\label{fig:qp}
\vspace{-5mm}
\end{figure}

Since the GAN architecture and the multi-resolution STFT loss of PWG have shown the effectiveness for training a speech generator, the proposed QPPWG inherits the discriminator and the $L_{\mathrm{sp}}$ regularizer from PWG and focuses on improving the generator using a QP structure. The main modules of the QP structure are PDCNN components and a cascaded architecture, which are inspired by the pitch filtering and the cascaded recursive structure of the code-excited linear prediction (CELP) codec~\cite{celp}.

\subsection{Pitch-dependent dilated convolution}

As shown in Fig.~\ref{fig:qp}, there are gaps between the inputs of a DCNN kernel, and the length of each gap is a predefined hyperparameter called a dilation size (rate). PDCNN is an extension of DCNN, and its dilation size is pitch-dependent and dynamically changed according to the input pitch. Specifically, a dilated convolution with a kernel size three is formulated as
\begin{align}
 \boldsymbol{y}_{t}^{(\mathrm{o})}
=\boldsymbol{W}^{(\mathrm{c})}\ast\boldsymbol{y}_{t}^{(\mathrm{i})}
+\boldsymbol{W}^{(\mathrm{p})}\ast\boldsymbol{y}_{t-d}^{(\mathrm{i})}
+\boldsymbol{W}^{(\mathrm{f})}\ast\boldsymbol{y}_{t+d}^{(\mathrm{i})},
\end{align}
where $\boldsymbol{y}^{(\mathrm{i})}$ and $\boldsymbol{y}^{(\mathrm{o})}$ are the input and output of the DCNN layer. $\boldsymbol{W}^{(\mathrm{c})}$, $\boldsymbol{W}^{(\mathrm{p})}$ and $\boldsymbol{W}^{(\mathrm{f})}$ are the trainable $1 \times 1$ convolution filters of current, previous, and following samples, respectively. $\ast$ is the convolution operator, and the dilation size $d$ of DCNN is a time-invariant constant. By contrast, PDCNN adopts a pitch-dependent dilated factor $E_{t}$ to dynamically change the dilation size $d^{\prime}$ in each time step $t$ as
\begin{align}
d^{\prime}=E_{t} \times d.
\end{align}
The dilated factor $E_{t}$ is derived from
\begin{align}
E_{t}=F_{s}/(F_{0,t}\times a),
\end{align}
where $F_{s}$ is the sampling rate, $F_{0}$ is the fundamental frequency, and $a$ is a hyperparameter called a {\it dense factor}, which indicates the number of samples in one cycle taken as the PDCNN inputs for each time step. The higher the {\it dense factor}, the lower the sparsity of PDCNN. For simplicity,  the $d^{\prime}$ of each time step is rounded. For the unvoiced segments, we find that the models with interpolated $F_{0}$ outperform the models setting $d^{\prime}$ to one, so our implementation~\cite{github} adopts interpolated $F_{0}$. In conclusion, PDCNN introduces the pitch information to the network, makes each sample have a pitch-dependent {\it receptive field} size, and efficiently enlarges {\it receptive fields}.

\begin{figure}[t]
\begin{center}
\includegraphics[width=0.9\columnwidth]{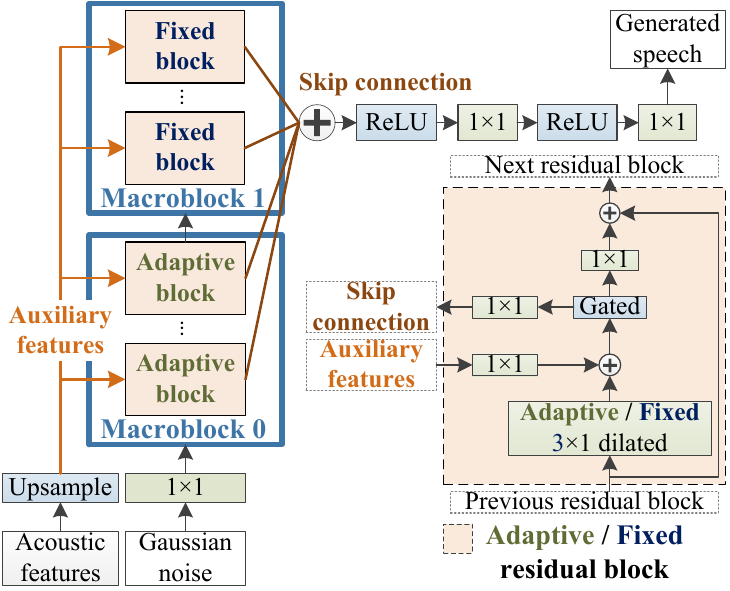}
\caption{Architecture of QPPWG generator.}
\label{fig:qppwg}
\end{center}
\vspace{-10mm}
\end{figure}

\subsection{Generator of QPPWG}

The architecture of the QPPWG generator is shown in Fig.~\ref{fig:qppwg}, and it is similar to the WN-like PWG generator. The main difference is the hierarchical architecture of the stacked residual blocks. Specifically, QPPWG includes two cascaded macroblocks with different types of residual blocks while PWG only contains one type of residual blocks. The first macroblock of QPPWG consists of stacked adaptive blocks with PDCNN layers to model the periodic components with long-term correlations, and the second macroblock of QPPWG consists of stacked fixed blocks with DCNN layers to model the aperiodic components with short-term correlations.

\section{Experiments}

\subsection{Model descriptions}

In this paper, the QPPWG models with two different orders of macroblocks and the PWG models with two different numbers of residual blocks were evaluated. Specifically, the QPPWG model, whose first macroblock included 10 adaptive blocks with 2 cycles of the dilation size expansions (B$_\mathrm{A}$10C2) and the following macroblock included 10 fixed blocks with one cycle (B$_\mathrm{F}$10C1), was denoted as QPPWG${af}$ and the one with a reverse macroblock order was QPPWG${fa}$. The generator of vanilla PWG (PWG\_30) contained 30 fixed blocks with 3 cycles (B$_\mathrm{F}$30C3) and the compact PWG (PWG\_20) contained 20 fixed blocks with 2 cycles (B$_\mathrm{F}$20C2). The channel and kernel sizes of the non-causal convolution of these generators were 64 and three. As shown in Table~\ref{tb:model_size}, the generator size of the proposed QPPWG is around 70~\% of that of the vanilla PWG because of the less stacked residual blocks. However, because the time-variant mechanism degrades the parallelism of CNNs, the real-time factor (RTF) of QPPWG generation with a Titan V GPU is around 0.020, which is higher than 0.016 of PWG\_30 and 0.011 of PWG\_20. Moreover, these four models adopted the the same discriminator architecture including 10 non-causal DCNN layers with 64 convolution channels, three kernels, and LeakyReLU ($\alpha$ is 0.2) activation functions, and the number of the discriminator parameters was around 0.1~M.

\begin{table}[t]
\caption{Numbers of generator parameters.}
\vspace{-3.5mm}
\label{tb:model_size}
\fontsize{8pt}{9.6pt}
\selectfont
{%
\begin{tabularx}{\columnwidth}{@{}p{1cm}YYYY@{}}
\toprule
         & PWG\_30    & PWG\_20    & QPPWG${af}$ & QPPWG${fa}$ \\ \midrule
Macro 0  & B$_\mathrm{F}$30C3      & B$_\mathrm{F}$20C2     
         & B$_\mathrm{A}$10C2      & B$_\mathrm{F}$10C1      \\
Macro 1  & -          & -          & B$_\mathrm{F}$10C1 & B$_\mathrm{A}$10C2 \\ \midrule
Size (M) & 1.16       & 0.78       & 0.79        & 0.79        \\ \bottomrule
\end{tabularx}%
}
\vspace{1mm}
\caption{Objective evaluation results.}
\vspace{-3.5mm}
\label{tb:objective}
\fontsize{8pt}{9.6pt}
\selectfont
{%
\begin{tabularx}{\columnwidth}{@{}p{1.5cm}YYYYY@{}}
\toprule
Vocoder & WD 
        & \multicolumn{2}{c}{PWG} 
        & \multicolumn{2}{c}{QPPWG}         \\
Blocks  & - &30 & 20 & $af$ & $fa$ \\ \midrule
        & \multicolumn{5}{c}{\cellcolor[HTML]{F2F2F2}RMSE of log $F_0$} \\
$1\times F_0$    & \cellcolor[HTML]{D9D9D9}0.10 & 0.12 & 0.15 & \textbf{0.11} & \textbf{0.11} \\
$1/2\times F_0$  & \cellcolor[HTML]{D9D9D9}0.14 & 0.27 & 0.32 & \textbf{0.19} & 0.20 \\
$2\times F_0$    & \cellcolor[HTML]{D9D9D9}0.10 & 0.15 & 0.15 & \textbf{0.11} & \textbf{0.11} \\ 
\textbf{Average} & \cellcolor[HTML]{D9D9D9}0.11 & 0.18 & 0.21 & \textbf{0.14} & 0.14 \\
\midrule
        & \multicolumn{5}{c}{\cellcolor[HTML]{F2F2F2}MCD (dB)} \\
$1\times F_0$    & \cellcolor[HTML]{D9D9D9}2.58 & \textbf{3.69} & 3.74 & 3.80 & 4.54 \\
$1/2\times F_0$  & \cellcolor[HTML]{D9D9D9}3.89 & 4.47 & \textbf{4.39} & 4.52 & 5.18 \\
$2\times F_0$    & \cellcolor[HTML]{D9D9D9}3.79 & 5.24 & 5.06 & \textbf{4.92} & 5.61 \\ 
\textbf{Average} & \cellcolor[HTML]{D9D9D9}3.42 & 4.46 & \textbf{4.40} & 4.41 & 5.11 \\ 
\bottomrule
\end{tabularx}%
}
\vspace{-8mm}
\end{table}

\subsection{Experimental settings}

All NN-based vocoders were trained in a multi-speaker manner. The training corpus consisted of 2200 utterances of the ``slt'' and ``bdl'' speakers of CMU-ARCTIC~\cite{arctic} and 852 utterances of all speakers of Voice Conversion Challenge 2018 (VCC2018)~\cite{vcc2018}. The total data length was around 2.5 hours. The testing corpus was the SPOKE set, which consisted of two male and two female speakers, of VCC2018 corpus, and the number of testing utterances of each speaker was 35. All speech data were set to a sampling rate of 22,050~Hz and a 16-bit resolution.

The auxiliary features of these speech generation models consisted of one-dimensional continuous $F_{0}$, one-dimensional unvoiced/voiced binary code ($U/V$), 35-dimensional mel-cepstrum ($mcep$), and two-dimensional coded aperiodicity ($codeap$). The WD vocoder was first adopted to extract one- dimensional $F_{0}$ and 513-dimensional spectral feature ($sp$) and aperiodicity ($ap$) with a frameshift of 5 ms. $F_{0}$ was interpolated to the continuous $F_{0}$ and converted to the $U/V$, $ap$ was coded into the $codeap$, and $sp$ was parameterized into the $mcep$. To simulate unseen data, the continuous $F_{0}$ was scaled by the ratios of 1/2 and 2 while keeping other features the same. Moreover, the dilated factor $E_{t}$ of QPPWG was calculated with the continuous $F_{0}$ because of the higher speech quality, and the {\it dense factor} of QPPWG was empirically set to four.

The models were trained with a RAdam optimizer~\cite{radam} ($\epsilon=1e^{-6}$) with 400~k iterations. For the stability, the generators of the models were trained with only multi-resolution STFT losses, which were calculated on the basis of three different FFT sizes
(1024 / 2048 / 512), frameshifts (120 / 240 / 50), and frame lengths (600 / 1200 / 240), for the first 100~k iterations and then jointly trained with the discriminators for the following 300~k iterations. The balance weight $\lambda_{\mathrm{adv}}$ of $L_{\mathrm{adv}}$ was set to
4.0. The learning rates, which decayed 50~\% every 200~k iterations, of the generators were $1e^{-4}$ and the discriminators were $5e^{-5}$. The minibatch size was six and the batch length was 25,520 samples.

\begin{figure}[t]
\begin{center}
 \includegraphics[width=0.95\columnwidth]{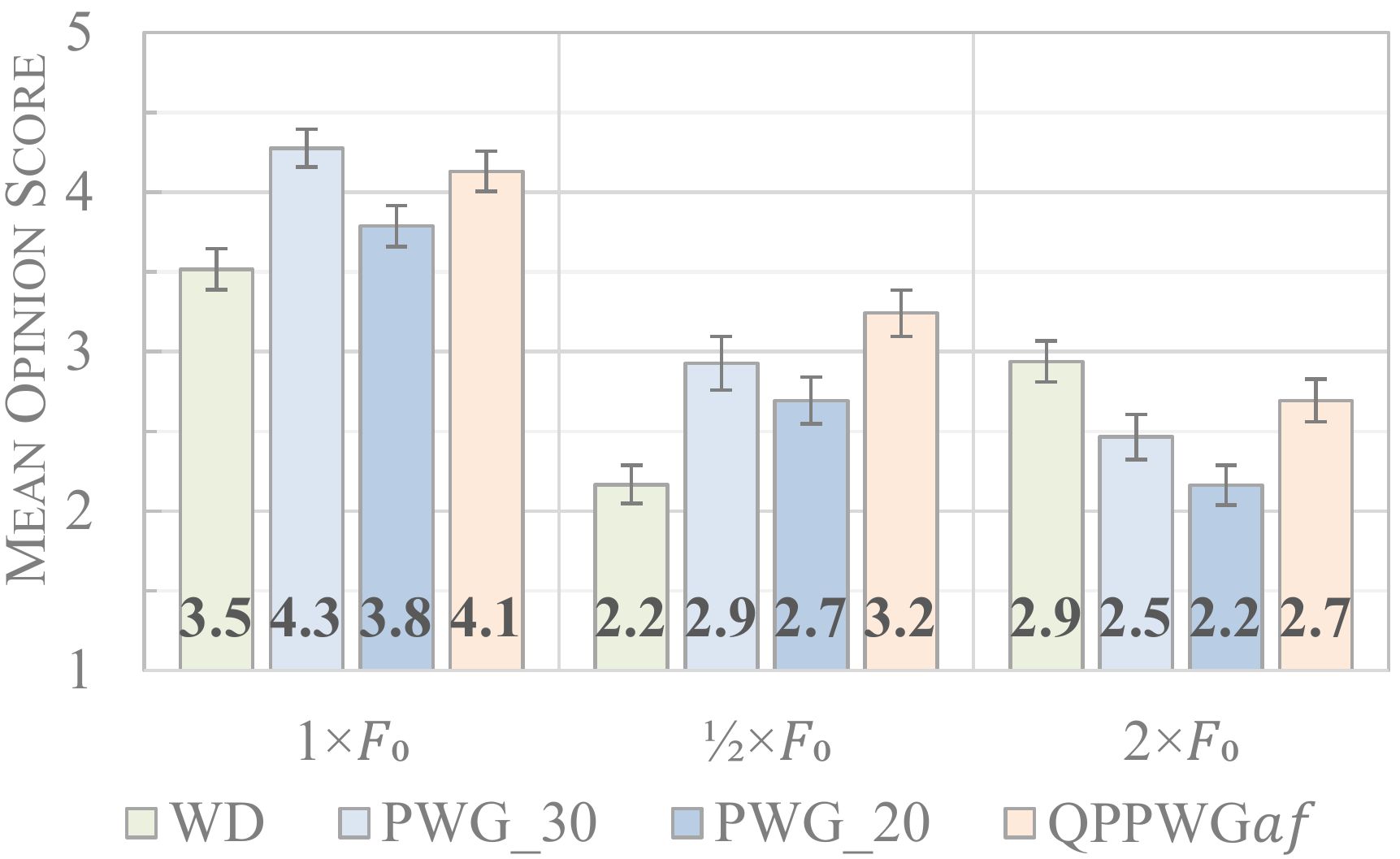}
\vspace{-3mm}
\caption{MOS results of speech quality with 95~\% CI.}
\label{fig:mos}

{\ }

\includegraphics[width=0.95\columnwidth]{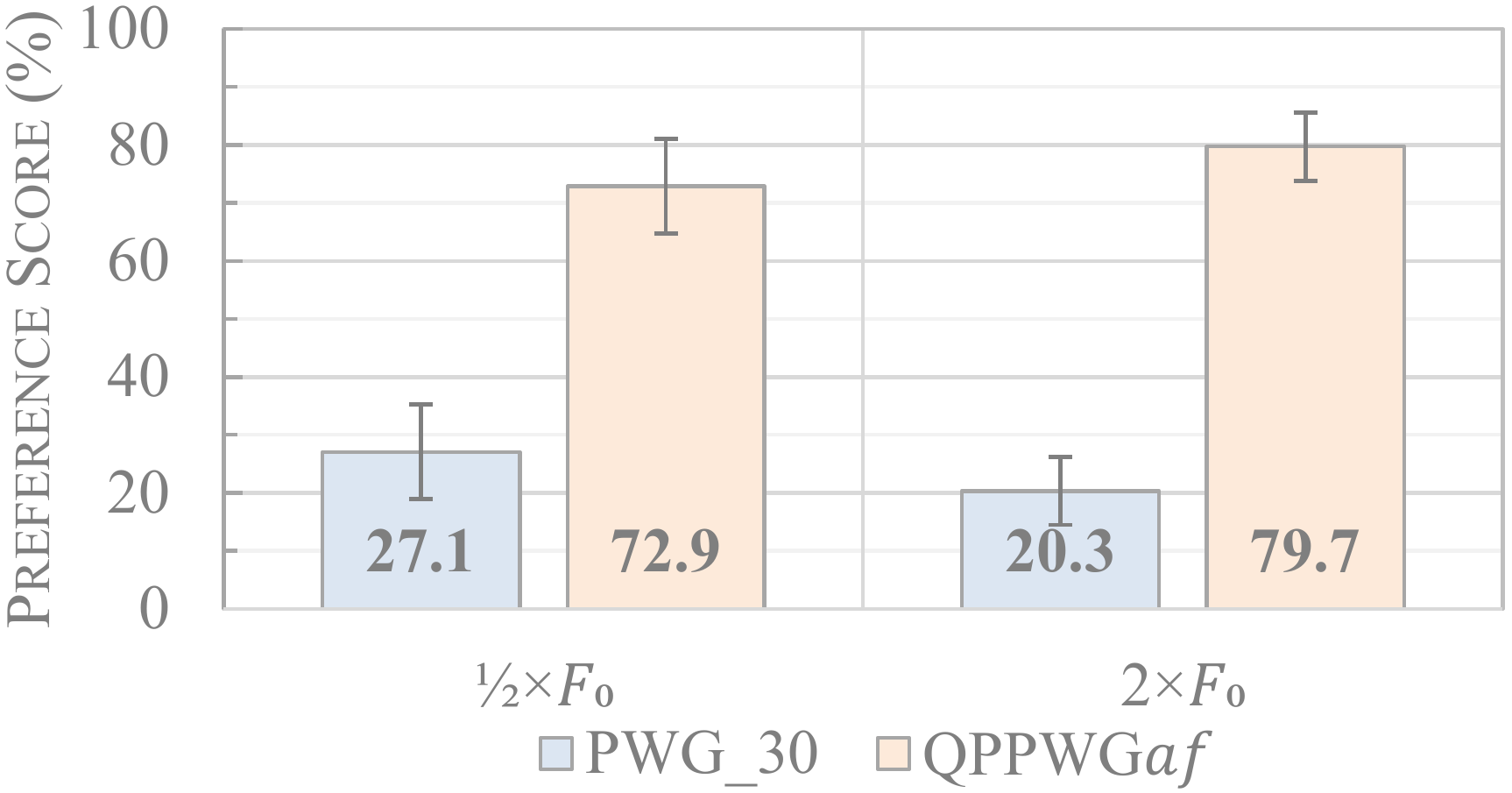}
\vspace{-3mm}
\caption{ABX results of pitch accuracy with 95~\% CI.}
\label{fig:abx}
\end{center}
\vspace{-10mm}
\end{figure}

\subsection{Objective evaluations}

Root mean square error (RMSE) of log $F_{0}$ and mel-cepstral distortion (MCD) were adopted to the objective evaluations. Both measurements were calculated using the auxiliary features and the features extracted from the generated speech. As shown in Table~\ref{tb:objective}, the proposed QPPWG vocoders achieve markedly higher $F_{0}$ accuracy than the PWG vocoders when conditioned on the scaled $F_{0}$, and it confirms the effectiveness of the proposed QP structure. The results also show that QPPWG${af}$ achieves a comparable spectral prediction accuracy as the vanilla and compact PWGs. Moreover, QPPWG${af}$ achieving lower MCDs than QPPWG${fa}$ implies that modeling long-term correlations first get a better overall spectral structure. In conclusion, the proposed QP structure improves the accuracy of pitch modeling of the PWG vocoder and efficiently extends the {\it receptive field} length. The QPPWG with an adaptive to fixed order outperforms the QPPWG with a reverse order.

\subsection{Subjective evaluations}

The subjective evaluation set consisted of 960 selected utterances of four testing speakers, four vocoders (WD, PWG\_30, PWG\_20, and QPPWG${af}$), and three $F_{0}$ scaled ratios (1, 1/2, and 2). For each speaker, vocoder, and $F_{0}$ ratio, we randomly selected 20 utterances from the 35 testing utterances for both mean opinion score (MOS) and ABX tests. Specifically, the speech quality of each utterance was evaluated by listeners assigning MOSs (1--5). The higher the MOS, the better the speech quality. For the ABX test, every time listeners compared two testing utterances with one reference to pick up the utterance whose pitch contour was more consistent with that of the reference. The WD-generated utterances were taken as the references, and the pitch accuracies of the QPPWG${af}$-generated utterances with 1/2 and 2 $F_{0}$ inputs were compared with that of the PWG\_30-generated utterances. Eight listeners involved in both tests and each utterance was evaluated by at least two listeners. Most listeners were audio-related researchers.

As shown in Fig.~\ref{fig:mos}, the QPPWG${af}$ vocoder markedly outperforms the same sized PWG\_20 vocoder for all $F_{0}$ inputs. Even compared with PWG\_30, QPPWG${af}$ still achieves higher speech qualities for the scaled $F_{0}$ inputs and a comparable speech quality for the unchanged $F_{0}$ input. In addition, the results of Fig.~\ref{fig:abx} show the perceptible differences of the pitch accuracies between the
QPPWG${af}$ and PWG\_30 vocoders with scaled $F_{0}$ inputs. In conclusion, introducing the pitch information to the PWG model by the QP structure markedly improves the pitch and speech modeling capabilities of the PWG vocoder, which results in compact model size and better pitch controllability of the QPPWG voder.

\begin{figure}[t]
\fontsize{8pt}{9pt}
\selectfont
{%
\begin{tabularx}{1\columnwidth}{@{}YY@{}}
\toprule
  \centering\arraybackslash{QPPWG$af$} 
& \centering\arraybackslash{QPPWG${fa}$} \\ \midrule
  \multicolumn{2}{c}{Intermediate cumulative outputs of 1--10 blocks} \\ 
  \includegraphics[width=0.44\columnwidth]{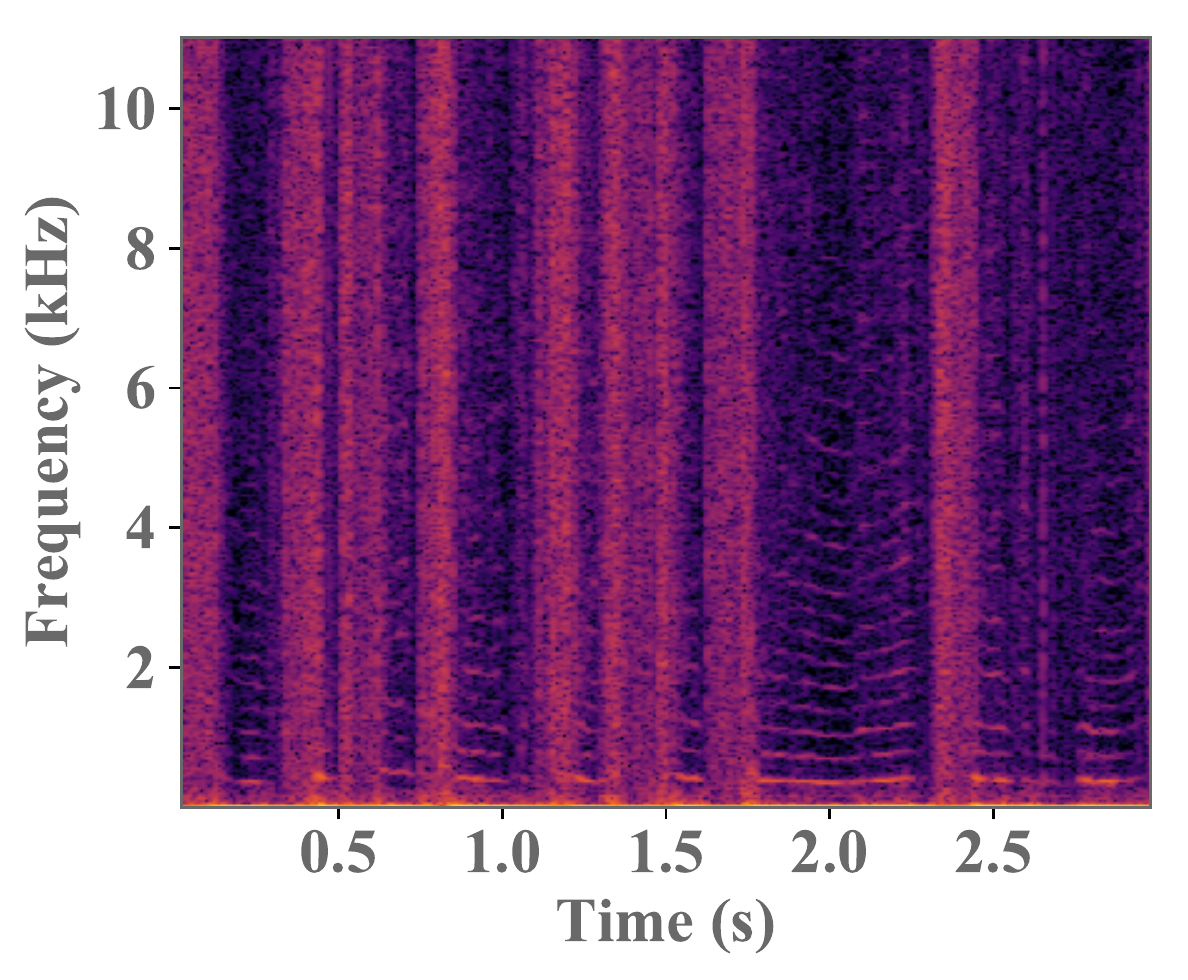}
& \includegraphics[width=0.44\columnwidth]{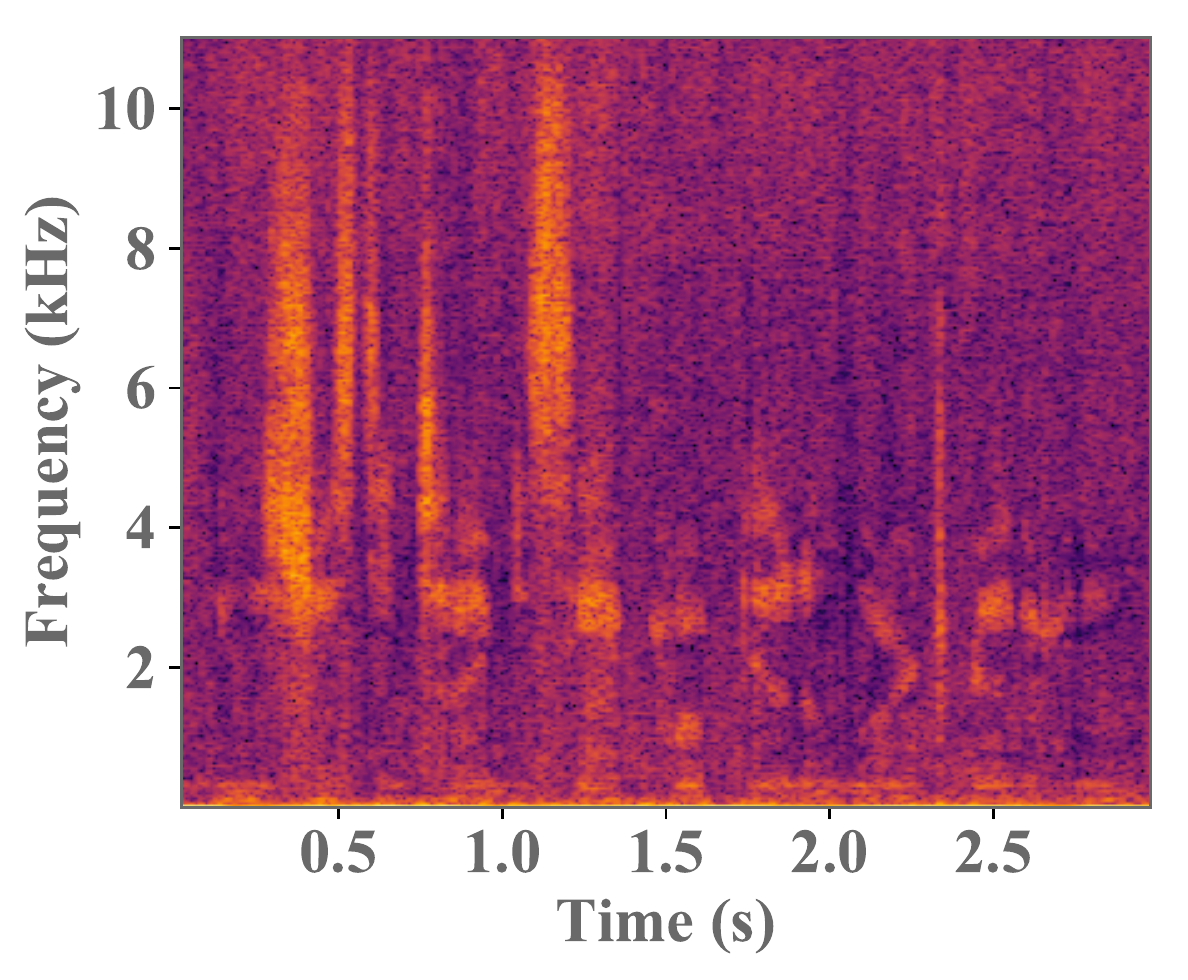} \\
  \multicolumn{2}{c}{Cumulative outputs of 1--20 blocks} \\ 
  \includegraphics[width=0.44\columnwidth]{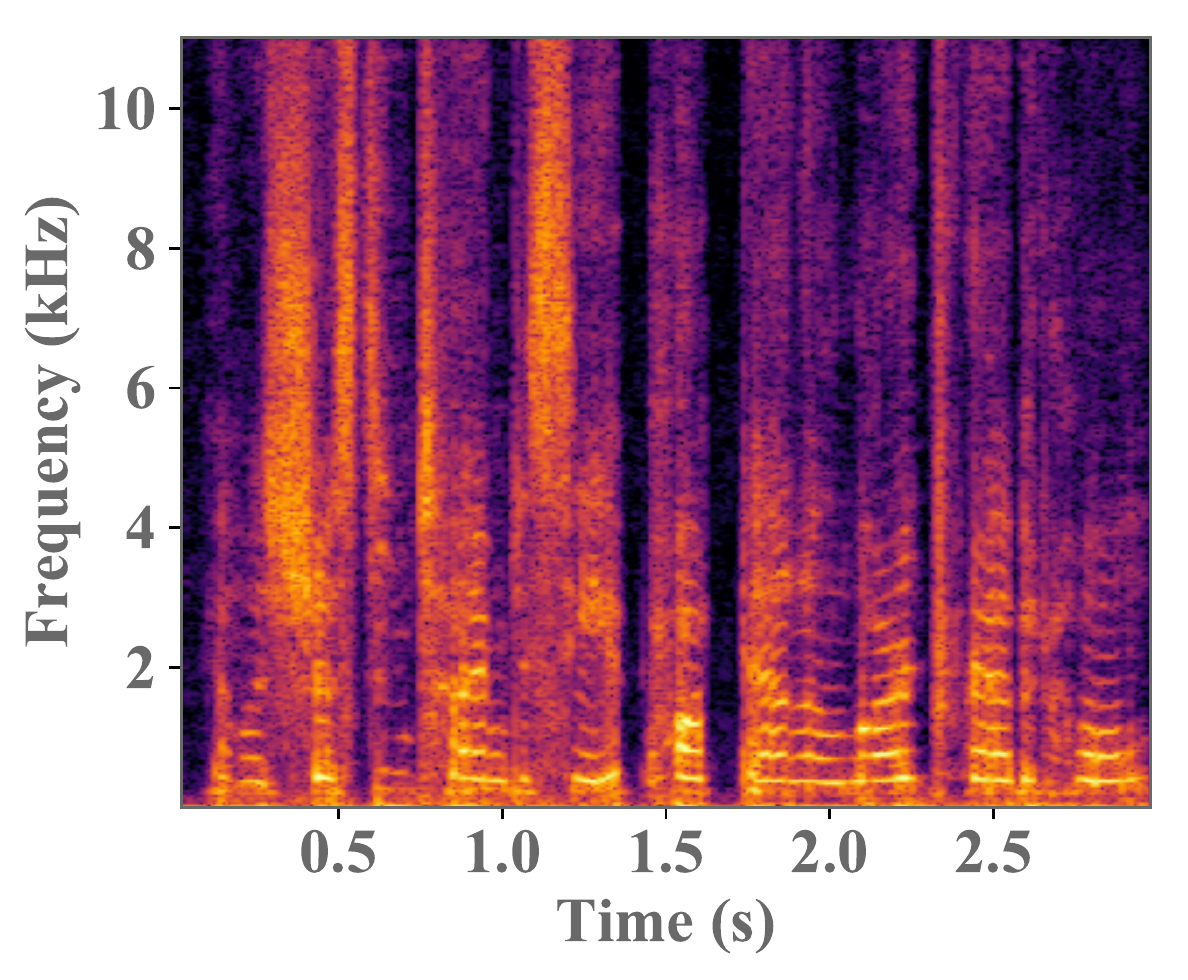}
& \includegraphics[width=0.44\columnwidth]{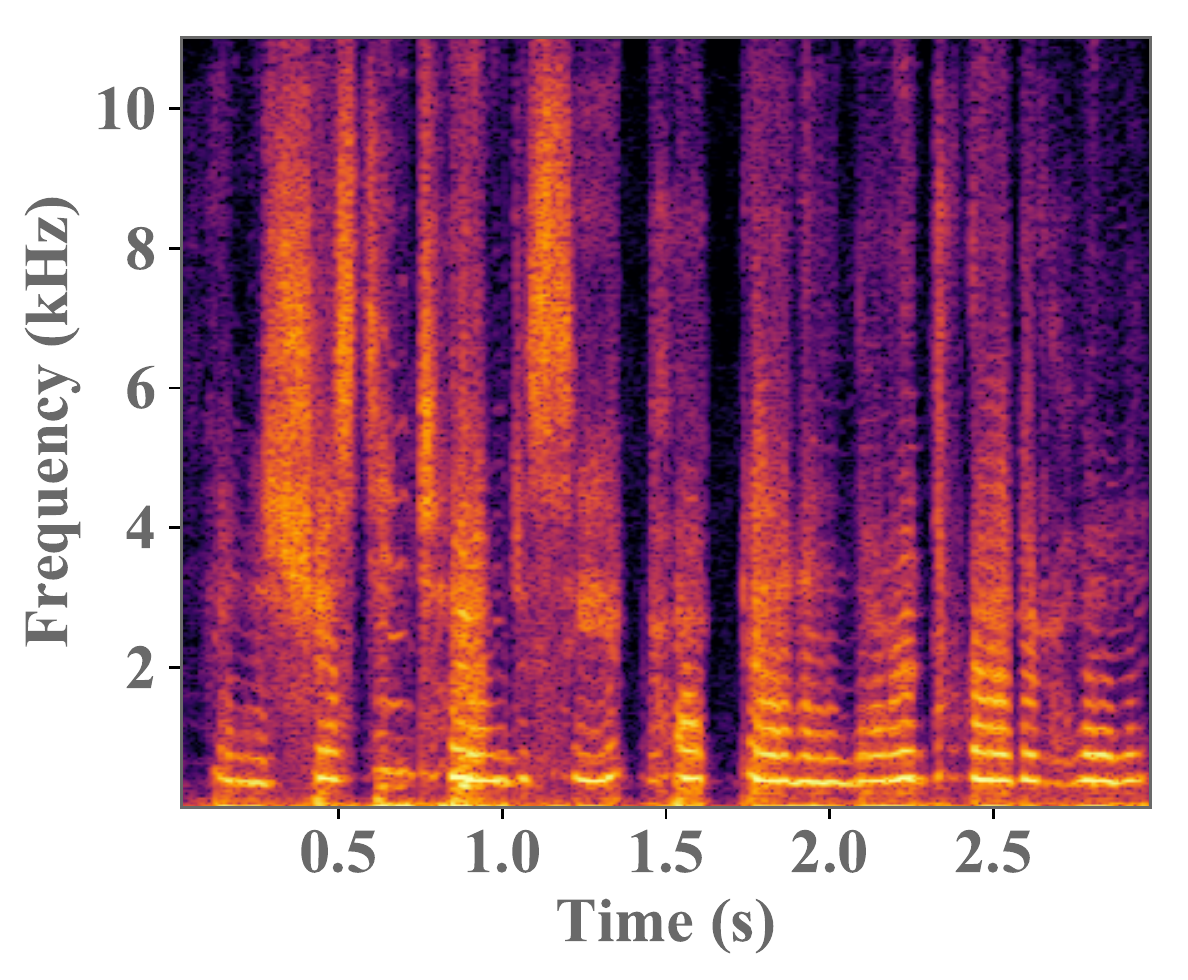} \\
\bottomrule
\end{tabularx}%
}
\vspace{-2mm}
\caption{Spectra of cumulative outputs.}
\label{fig:cumulative}
\vspace{-5mm}
\end{figure}

\subsection{Discussion}

Since the model capacity is highly related to the {\it receptive field}
length~\cite{qpnet_2019,qpnet_2020}, and the length of PWG\_30 is 6139 samples
($2^{0}+\cdots +2^{9}=1023$ with three cycles and two sides plus one),
QPPWG attains a longer {\it effective receptive field} length around 3,000--16,000
samples. Specifically, the size is 2047 of the B$_\mathrm{F}$10C1 and
124$\times E_{T}$ ($2^{0}+\cdots +2^{4}=31$ with two cycles and two sides)
of the B$_\mathrm{A}$10C2, and the $E_{T}$ is around 11--110 of the
500--50 Hz pitches when the {\it dense factor} is four.

Moreover, as the intermediate cumulative outputs shown in
Fig.~\ref{fig:cumulative}, the first ten adaptive blocks of QPPWG${af}$
focus on modeling the pitch and harmonic components, which have
long-term correlations, while the first ten fixed blocks of QPPWG${fa}$
focus on modeling the non-harmonic components, which have short-term
correlations. The results confirm our assumptions of the QP structure,
and the behavior, which is similar to the harmonic plus noise
model~\cite{hpn_1,hpn_2}, of QPPWG is more tractable and interpretable
than that of vanilla PWG. More details and demo samples can be found on
our website~\cite{demo}.

\section{Conclusions}

In this paper, we integrate a fast and compact PWG vocoder with a QP structure to improve its pitch controllability. The proposed QPPWG vocoder reduces the model size and achieves higher speech quality and pitch accuracy than the PWG vocoder when the input $F_{0}$ sequence is scaled. In conclusion, the QPPWG vocoder is more in line with the definition of a vocoder, which attains acoustic controllability.

\section{Acknowledgments}

This work was supported in part by JST CREST Grant Number JPMJCR19A3. The initial investigation in this study was performed while Y.-C. Wu was interning at NICT.

\bibliography{mybib}

\begin{thebibliography}{10}
\providecommand{\url}[1]{#1}
\csname url@samestyle\endcsname
\providecommand{\newblock}{\relax}
\providecommand{\bibinfo}[2]{#2}
\providecommand{\BIBentrySTDinterwordspacing}{\spaceskip=0pt\relax}
\providecommand{\BIBentryALTinterwordstretchfactor}{4}
\providecommand{\BIBentryALTinterwordspacing}{\spaceskip=\fontdimen2\font plus
\BIBentryALTinterwordstretchfactor\fontdimen3\font minus
  \fontdimen4\font\relax}
\providecommand{\BIBforeignlanguage}[2]{{%
\expandafter\ifx\csname l@#1\endcsname\relax
\typeout{** WARNING: IEEEtran.bst: No hyphenation pattern has been}%
\typeout{** loaded for the language `#1'. Using the pattern for}%
\typeout{** the default language instead.}%
\else
\language=\csname l@#1\endcsname
\fi
#2}}
\providecommand{\BIBdecl}{\relax}
\BIBdecl

\bibitem{vocoder_1939}
H.~Dudley, ``The vocoder,'' \emph{Bell Labs Record}, vol.~18, no.~4, pp.
  122--126, 1939.

\bibitem{vocoder_1966}
M.~R. Schroeder, ``Vocoders: Analysis and synthesis of speech,'' \emph{Proc.
  IEEE}, vol.~54, no.~5, pp. 720--734, 1966.

\bibitem{straight}
H.~Kawahara, I.~Masuda-Katsuse, and A.~De~Cheveigne, ``Restructuring speech
  representations using a pitch-adaptive time--frequency smoothing and an
  instantaneous-frequency-based f0 extraction: Possible role of a repetitive
  structure in sounds,'' \emph{Speech communication}, vol.~27, no. 3-4, pp.
  187--207, 1999.

\bibitem{world}
M.~Morise, F.~Yokomori, and K.~Ozawa, ``World: a vocoder-based high-quality
  speech synthesis system for real-time applications,'' \emph{IEICE
  TRANSACTIONS on Information and Systems}, vol.~99, no.~7, pp. 1877--1884,
  2016.

\bibitem{source_filter}
R.~McAulay and T.~Quatieri, ``Speech analysis/synthesis based on a sinusoidal
  representation,'' \emph{IEEE Transactions on Acoustics, Speech, and Signal
  Processing}, vol.~34, no.~4, pp. 744--754, 1986.

\bibitem{samplernn}
S.~Mehri, K.~Kumar, I.~Gulrajani, R.~Kumar, S.~Jain, J.~Sotelo, A.~Courville,
  and Y.~Bengio, ``Sample{RNN}: {A}n unconditional end-to-end neural audio
  generation model,'' in \emph{Proc. ICLR}, Apr. 2017.

\bibitem{wavenet}
A.~van~den Oord, S.~Dieleman, H.~Zen, K.~Simonyan, O.~Vinyals, A.~Graves,
  N.~Kalchbrenner, A.~Senior, and K.~Kavukcuoglu, ``Wave{N}et: A generative
  model for raw audio,'' in \emph{Proc. SSW9}, Sept. 2016, p. 125.

\bibitem{sd_wn_vocoder}
A.~Tamamori, T.~Hayashi, K.~Kobayashi, K.~Takeda, and T.~Toda,
  ``Speaker-dependent {W}ave{N}et vocoder,'' in \emph{Proc. Interspeech}, Aug.
  2017, pp. 1118--1122.

\bibitem{si_wn_vocoder}
T.~Hayashi, A.~Tamamori, K.~Kobayashi, K.~Takeda, and T.~Toda, ``An
  investigation of multi-speaker training for {W}ave{N}et vocoder,'' in
  \emph{Proc. ASRU}, Dec. 2017, pp. 712--718.

\bibitem{ns_wn_vocoder}
K.~Tachibana, T.~Toda, Y.~Shiga, and H.~Kawai, ``An investigation of noise
  shaping with perceptual weighting for wavenet-based speech generation,'' in
  \emph{2018 IEEE International Conference on Acoustics, Speech and Signal
  Processing (ICASSP)}.\hskip 1em plus 0.5em minus 0.4em\relax IEEE, 2018, pp.
  5664--5668.

\bibitem{srnn_vocoder}
Y.~Ai, H.-C. Wu, and Z.-H. Ling, ``Sample{RNN}-based neural vocoder for
  statistical parametric speech synthesis,'' in \emph{Proc. ICASSP}, Apr. 2018,
  pp. 5659--5663.

\bibitem{fftnet}
Z.~Jin, A.~Finkelstein, G.~J. Mysore, and J.~Lu, ``{FFTN}et: A real-time
  speaker-dependent neural vocoder,'' in \emph{Proc. ICASSP}, Apr. 2018, pp.
  2251--2255.

\bibitem{wavernn}
N.~Kalchbrenner, E.~Elsen, K.~Simonyan, S.~Noury, N.~Casagrande, E.~Lockhart,
  F.~Stimberg, A.~van~den Oord, S.~Dieleman, and K.~Kavukcuoglu, ``Efficient
  neural audio synthesis,'' in \emph{Proc. ICML}, July 2018, pp. 2415--2424.

\bibitem{lpcnet}
J.-M. Valin and J.~Skoglund, ``{LPCN}et: Improving neural speech synthesis
  through linear prediction,'' in \emph{Proc. ICASSP}, May 2019, pp.
  5826--7830.

\bibitem{pwn}
A.~van~den Oord, Y.~Li, I.~Babuschkin, K.~Simonyan, O.~Vinyals, K.~Kavukcuoglu,
  G.~van~den Driessche, E.~Lockhart, L.~C. Cobo, F.~Stimberg, N.~Casagrande,
  D.~Grewe, S.~Noury, S.~Dieleman, E.~Elsen, N.~Kalchbrenner, H.~Zen,
  A.~Graves, H.~King, T.~Walters, D.~Belov, and D.~Hassabis, ``Parallel
  {W}ave{N}et: Fast high-fidelity speech synthesis,'' in \emph{Proc. ICML},
  July 2018, pp. 3915--3923.

\bibitem{clarinet}
W.~Ping, K.~Peng, and J.~Chen, ``Clari{N}et: Parallel wave generation in
  end-to-end text-to-speech,'' in \emph{Proc. ICLR}, May 2019.

\bibitem{waveglow}
R.~Prenger, R.~Valle, and B.~Catanzaro, ``Wave{G}low: A flow-based generative
  network for speech synthesis,'' in \emph{Proc. ICASSP}, May 2019, pp.
  3617--3621.

\bibitem{flowavenet}
S.~Kim, S.-G. Lee, J.~Song, J.~Kim, and S.~Yoon, ``{F}lo{W}ave{N}et : A
  generative flow for raw audio,'' in \emph{Proc. ICML}, June 2019, pp.
  3370--3378.

\bibitem{waveffjord}
N.-Q. Wu and Z.-H. Ling, ``Wave{FFJORD}: {FFJORD}-based vocoder for statistical
  parametric speech synthesis,'' in \emph{Proc. ICASSP}, May 2020, pp.
  7214--7218.

\bibitem{gan}
I.~Goodfellow, J.~Pouget-Abadie, M.~Mirza, B.~Xu, D.~Warde-Farley, S.~Ozair,
  A.~Courville, and Y.~Bengio, ``Generative adversarial nets,'' in \emph{Proc.
  NIPS}, Dec. 2014, pp. 2672--2680.

\bibitem{pwg}
R.~Yamamoto, E.~Song, and J.-M. Kim, ``Parallel {W}ave{GAN}: A fast waveform
  generation model based on generative adversarial networks with
  multi-resolution spectrogram,'' in \emph{Proc. ICASSP}, May 2020, pp.
  6199--6203.

\bibitem{melgan}
K.~Kumar, R.~Kumar, T.~de~Boissiere, L.~Gestin, W.~Z. Teoh, J.~Sotelo,
  A.~de~Br\'{e}bisson, Y.~Bengio, and A.~C. Courville, ``Mel{GAN}: Generative
  adversarial networks for conditional waveform synthesis,'' in \emph{Proc.
  NeurIPS}, Dec. 2019, pp. 14\,910--14\,921.

\bibitem{gantts}
M.~Bi{\'n}kowski, J.~Donahue, S.~Dieleman, A.~Clark, E.~Elsen, N.~Casagrande,
  L.~C. Cobo, and K.~Simonyan, ``High fidelity speech synthesis with
  adversarial networks,'' in \emph{Proc. ICLR}, Apr. 2020.

\bibitem{nsf_2019}
X.~Wang, S.~Takaki, and J.~Yamagishi, ``Neural source-filter-based waveform
  model for statistical parametric speech synthesis,'' in \emph{Proc. ICASSP},
  May 2019, pp. 5916--5920.

\bibitem{nsf_2020}
X.~{Wang}, S.~{Takaki}, and J.~{Yamagishi}, ``Neural source-filter waveform
  models for statistical parametric speech synthesis,'' \emph{IEEE/ACM
  Transactions on Audio, Speech, and Language Processing}, vol.~28, pp.
  402--415, 2020.

\bibitem{pap_gan}
K.~Oura, K.~Nakamura, K.~Hashimoto, Y.~Nankaku, and K.~Tokuda, ``Deep neural
  network based real-time speech vocoder with periodic and aperiodic inputs,''
  in \emph{Proc. SSW10}, Sept. 2019, pp. 13--18.

\bibitem{qpnet_2019}
Y.-C. Wu, T.~Hayashi, P.~L. Tobing, K.~Kobayashi, and T.~Toda, ``Quasi-periodic
  {W}ave{N}et vocoder: A pitch dependent dilated convolution model for
  parametric speech generation,'' in \emph{Proc. Interspeech}, Sept. 2019, pp.
  196--200.

\bibitem{qpnet_2020}
Y.-C. {Wu}, T.~{Hayashi}, P.~L. {Tobing}, K.~{Kobayashi}, and T.~{Toda},
  ``Quasi-periodic {W}ave{N}et: An autoregressive raw waveform generative model
  with pitch-dependent dilated convolution neural network,'' \emph{IEEE/ACM
  Transactions on Audio, Speech, and Language Processing}, (submitted).

\bibitem{dcnn}
F.~Yu and K.~Vladlen, ``Multi-scale context aggregation by dilated
  convolutions,'' in \emph{Proc. ICLR}, May 2016.

\bibitem{leakyrelu}
A.~L. Maas, A.~Y. Hannun, and A.~Y. Ng, ``Rectifier nonlinearities improve
  neural network acoustic models,'' in \emph{Proc. ICML}, June 2013, pp. 3--11.

\bibitem{celp}
M.~Schroeder and B.~Atal, ``Code-excited linear prediction({CELP}):
  High-quality speech at very low bit rates,'' in \emph{Proc. ICASSP}, vol.~10,
  Apr. 1985, pp. 937--940.

\bibitem{github}
\BIBentryALTinterwordspacing
Y.-C. Wu, \emph{{QPPWG} repository}, Accessed: 2020. [Online]. Available:
  \url{https://github.com/bigpon/QPPWG}
\BIBentrySTDinterwordspacing

\bibitem{arctic}
J.~Kominek and A.~W. Black, ``The {CMU ARCTIC} speech databases for speech
  synthesis research,'' in \emph{Tech. Rep. CMU-LTI- 03-177}, 2003.

\bibitem{vcc2018}
J.~Lorenzo-Trueba, J.~Yamagishi, T.~Toda, D.~Saito, F.~Villavicencio,
  T.~Kinnunen, and Z.~Ling, ``The voice conversion challenge 2018: Promoting
  development of parallel and nonparallel methods,'' in \emph{Proc. Odyssey},
  June 2018, pp. 195--202.

\bibitem{radam}
L.~Liu, H.~Jiang, P.~He, W.~Chen, X.~Liu, J.~Gao, and J.~Han, ``On the variance
  of the adaptive learning rate and beyond,'' in \emph{Proc. ICLR}, Apr. 2020.

\bibitem{hpn_1}
Y.~Stylianou, O.~Capp\'{e}, and E.~Moulines, ``Continuous probabilistic
  transform for voice conversion,'' \emph{IEEE Trans. Speech Audio Process.},
  vol.~6, no.~2, pp. 131--142, 1998.

\bibitem{hpn_2}
Y.~Stylianou, ``Applying the harmonic plus noise model in concatenative speech
  synthesis,'' \emph{IEEE Trans. Speech Audio Process.}, vol.~9, no.~1, pp.
  21--29, 2001.

\bibitem{demo}
\BIBentryALTinterwordspacing
Y.-C. Wu, \emph{{QPPWG} demo}, Accessed: 2020. [Online]. Available:
  \url{https://bigpon.github.io/QuasiPeriodicParallelWaveGAN_demo/}
\BIBentrySTDinterwordspacing

\end{thebibliography}
\bibliographystyle{IEEEtran}

\end{document}